\documentclass[preprint,pre,showpacs,aps]{revtex4}

\usepackage{amssymb}
\usepackage{graphicx}
\usepackage{dsfont}

\begin{document}

\title{Transport properties through graphene grain boundaries: \\ strain effects versus lattice symmetry}

\author{V. Hung Nguyen$^{1,2,3}$\footnote{E-mail: viet-hung.nguyen@uclouvain.be}, Trinh X. Hoang$^3$, P. Dollfus$^2$, and J.-C. Charlier$^1$} \address{$^1$Institute of Condensed Matter and Nanosciences, Universit\'{e} catholique de Louvain, Chemin des \'{e}toiles 8, B-1348 Louvain-la-Neuve, Belgium \\ $^2$Institut d'Electronique Fondamentale, CNRS,	Univ. Paris Sud, Univercit\'{e} Paris-Saclay, 91405 Orsay, France \\ $^3$Center for Computational Physics, Institute of Physics, Vietnam Academy of Science and Technology, P.O. Box 429 Bo Ho, 10000 Hanoi, Vietnam}

\begin{abstract}
	As most materials available in macroscopic quantities, graphene appears in a polycrystalline form and thus contains grain boundaries. In the present work, the effect of uniaxial strain on the electronic transport properties through graphene grain boundaries is investigated using atomistic simulations. A systematic picture of the transport properties with respect to the strain and the lattice symmetry of graphene domains on both sides of the boundary is provided. In particular, it is shown that the strain engineering can be used to open a finite transport gap in all graphene systems where two domains exhibit different orientations. This gap value is found to depend on the strain magnitude, on the strain direction and on the lattice symmetry of graphene domains. By choosing appropriately the strain direction, a large transport gap of a few hundred meV can be achieved when applying a small strain of only a few percents. For a specific class of graphene grain boundary systems, the strain engineering can also be used to reduce the scattering on defects and hence to significantly enhance the conductance. With a large strain-induced gap, these graphene heterostructures are proposed to be possible candidates for highly sensitive strain sensors, flexible transistors and p--n junctions with a strong non-linear \textit{I--V} characteristics.  
\end{abstract}

\pacs{xx.xx.xx, yy.yy.yy, zz.zz.zz}
\maketitle

\section{Introduction}

Graphene-based nanostructures have attracted a great amount of attention from the scientific communities, in view of the numerous possibilities offered not only for fundamental science but also for practical applications \cite{ferr15}. This broad interest is basically motivated by the outstanding properties of graphene such as high carrier mobility, small spin-orbit coupling, superior
mechanical strength and stiffness, high electronic and thermal conductivities, optical transparency, ... Hence, graphene appears as a promising material for integration into a variety of electrical, spintronic and optical applications and also for improving the performance of flexible devices.

Regarding the production of large graphene area, the chemical vapor deposition (CVD) method is known to be the best technique as it is able of both high structural quality and wafer-scale growth \cite{zhan13}. Unfortunately, the transfer of graphene to insulating substrates remains a big challenge and considerable efforts are still needed to further improve growth process. In particular, CVD graphene is found to be polycrystalline in nature \cite{biro13,yazy14} and composed of many single-crystal graphene domains separated by grain boundaries (GB) of irregular shapes. The GBs appear as extended structural defects consisting of a random one-dimensional distribution of non-hexagonal rings (i.e., pentagon, heptagon, octagon, ...). These structural defects act as a source of intrinsic scattering and strongly affect the overall properties of wafer-scale graphene materials \cite{cumm14}. In particular, it has been shown that the intrinsic strength of graphene samples is strongly modified according to the type of grain boundaries \cite{ywei12,raso13,jhon13}. Additionally, the grain boundaries are an important source of scattering that limits drastically the carrier mobility in CVD graphene samples \cite{song12,tuan13} and consequently the performance of the corresponding electronic devices \cite{jime14}. Optical and thermal properties of graphene are also significantly affected by the presence of these structural defects \cite{zfei13,bala11,ylu012,sero13}. The effects of grain boundary on other transport phenomena such as magnetotransport and defect-induced magnetism have also been explored \cite{cumm142,dutt15}.

Besides its intrinsic defective nature, graphene also exhibits an important drawback: the lack of bandgap, which is a real problem for conventional applications in electronics \cite{schw10}. So far, many efforts in engineering bandgap in graphene have been made to solve this issue. For instance, techniques such as cutting 2D graphene sheets into narrow nanoribbons \cite{yhan07}, depositing graphene on hexagonal boron nitride substrate \cite{khar11}, doping graphene with nitrogen \cite{lher13}, applying an electric field perpendicularly to Bernal-stacking bilayer graphene \cite{zhan09}, creating graphene nanomeshes \cite{jbai10}, using hybrid graphene/hexagonal boron-nitride \cite{fior12} or vertical graphene channels \cite{brit12} are a few examples. Recently, it has also been suggested that the use of hetero-channels consisting of two graphene domains with different electronic structures could be a potential strategy \cite{chun14,hung15}. The idea is that by joining two semi-metallic graphene domains, a finite energy gap of conductance/current through the channel can be achieved if the Dirac cones of these domains are shifted from each other in the \textit{k}-space. Following this idea, a finite energy gap as large as a few hundred meV has been predicted in unstrained/strained graphene heterojunctions \cite{chun14} and in vertical channels made of misoriented graphene layers \cite{hung15} when strain is conveniently applied. Furthermore, the transport gap has been shown to be strongly dependent not only on the strain magnitude but also on both the strain direction and the lattice orientation. This idea of opening a finite transport gap due to the mismatch between Dirac cones has been also explored in hetero-structures based on the vertical stacking of two different Dirac materials (graphene, silicene, germanene, ...) \cite{wang15}. Given the superior ability to sustain a large strain of over 20$\%$ \cite{kim009}, graphene appears as a promising material for flexible electronic devices \cite{shar03}. In this regard, the studies in \cite{chun14,hung15} have suggested a new route to enlarge the possibility of using graphene in strain sensors and flexible transistors. It is worth noting that strain engineering has been also demonstrated to be an alternative/efficient approach to modulate electrical, optical and magnetic properties of several graphene nanostructures \cite{pere09,stap01,stap02,stap03,stap04,stap05,stap06,stap07}.

In polycrystalline graphene, graphene domains with different orientations are frequently surrounding the grain boundary. This naturally offers the possibility of designing heterostructures with similar properties as those studied in \cite{chun14,hung15}. Unquestionably, a finite transport gap can be achieved in presence of a mismatch between the bandstructures of these two graphene domains. Indeed, this has been demonstrated in \cite{yazy10}, where a large energy gap of about 1 eV is achieved a graphene system containing one armchair oriented and one zigzag oriented domains surrounding both sides of the grain boundary. However, in graphene systems made of two commensurable lattices, the gap remains zero. In the study  \cite{kuma12}, strain engineering has been demonstrated to be efficient to modulate the electronic transport through such the graphene grain boundaries. In particular, the transport gap in the system of armchair and zigzag domains can be strongly modified by strain. In the systems of commensurable graphene lattices, a finite gap can be achieved when strain is applied if the two domains are asymmetric with respect to the grain boundary. However, to obtain a large modulation of transport gap and hence of conductance, a large strain of about 15-20$\%$ is required. Moreover, the strain engineering can not change the semi-metallic behavior of the symmetrical systems. Note that in the study \cite{kuma12}, only one strain direction (perpendicularly to the grain boundary) was investigated. Because, as suggested in similar graphene heterosystems \cite{chun14,hung15}, the transport gap strongly depends not only on the strain magnitude but also on the strain direction, the strain direction in \cite{kuma12} may not be the optimal one to achieve a large gap. In addition, since the transport gap is essentially due to the mismatch between the electronic structures of two graphene domains, we anticipate that the lattice symmetry of these two domains should play an important role, which has not been properly investigated yet.

In this context, we aim to investigate systematically the effects of uniaxial strain on the transport properties of graphene-based systems containing a single grain boundary. In particular, we first focus on the dependence of transport gap on (i) the amplitude and direction of strain and (ii) the lattice symmetries of graphene domains, i.e. their orientations, supercells, their relative commensurability .... Based on this, we explore the possibilities of strongly modulating/opening the transport gap in all the cases where the graphene domains have different orientations and for small strains of only a few percents. We also report on a new property of the scattering on defects in graphene grain boundary systems when strain is applied. Finally, we discuss the possible applications of this type of graphene heterostructures.

\section{Model and methodology}

The investigated graphene-based systems containing a single grain boundary that presents periodic pentagon-heptagon pairs along the Oy direction are illustrated as in Fig.1. In this work, we consider the electronic transport along the Ox direction, i.e., perpendicularly to the grain boundary. A uniaxial strain applied to the system is characterized by its magnitude $\sigma$ and its direction $\theta$ with respect to the transport direction (see the inset on the right of Fig.1).

In most cases, graphene domains surrounding the boundary exhibit different crystallographic orientations. More specifically, we distinguish two typical cases for these domains depending on their lattice relative symmetries (see Fig.1). In both cases, the domains are determined by rotating the original (zigzag- or armchair-oriented) lattice by an angle $\phi_{rot}$. This angle is actually the angle between translation vectors $\mathbf{{L_y}}$ of rotated lattice and $\mathbf{{L_y^0}}$ of original one (see the images in the top of Fig.1) and determined as $\cos \phi_{rot} = \frac{{\mathbf{{L_y}} \cdot \mathbf{L_y^0}}}{{\|\mathbf{L_y}\| \|\mathbf{L_y^0}\|}}$. These vectors are simply given by $\mathbf{L_y^0} = \mathbf{a_1} + \mathbf{a_2}$ while $\mathbf{L_y} = p \mathbf{a_1} + q \mathbf{a_2}$. Before the rotation, $\mathbf{a_{1,2}} = (\pm \sqrt{3},3)a_0/2$ and $(\pm 3,\sqrt{3})a_0/2$ in the zigzag- and armchair-oriented lattices, respectively, where $a_0$ is the in-plane \textit{C-C} bond length ($= 1.42$ \AA{°}) in pristine graphene. Hereafter, these zigzag- and armchair-oriented lattices are identified with the denominations ZZ$p,q$ and AM$p,q$, respectively. Accordingly, we determine also the translational vectors along the Ox axis as $\mathbf{L_x} = n \mathbf{a_1} + m \mathbf{a_2}$ that satisfies the condition $\mathbf{L}_x\cdot\mathbf{L}_y = 0$, i.e., $\frac{{{n}}}{{{m}}} =  - \frac{{2{q} + {p}}}{{2{p} + {q}}}$ for ZZ$_{p,q}$ and $\frac{{{n}}}{{{m}}} = - \frac{{2{q} - {p}}}{{2{p} - {q}}}$ for AM$_{p,q}$ lattices. We also distinguish two types of grain boundary systems where the left and right graphene domains are commensurable or incommensurable. In the commensurable cases, one can find a common periodicity for the two graphene domains by extending the supercell appropriately from the original lattices, which is not the case of the incommensurable lattices. Hence, the translational length $L_y$ is the same for two domains in the commensurable systems while there is a mismatch between their translational lengths in the incommensurable cases. In the incommensurable systems studied in the present work, the mismatch mentioned above is small (i.e., $\lesssim 3 \%$) allowing the calculations to be performed using the average translational length as the size for the periodic cells along the Oy axis \cite{yazy10}.

Each graphene-based system was relaxed using classical molecular dynamics to minimize its energy determined by optimized Tersoff potentials \cite{lind10}. This empirical potential model is shown to accurately describe the structural properties and to provide the accurate calculations of thermal/mechanical properties in pristine and defective graphene including grain boundaries \cite{ylu012,sero13,bren01,bren02,bren03,bren04,bren05,bren06}. In order to compute the electronic transport quantities, we then employed a non-equilibrium Green's function formalism to solve the $\pi$-orbital tight-binding model. It is worth noting that the agreement between tight-binding calculation and density functional theory in graphene GBs has been demonstrated in \cite{yazy10}. It has to be also noted that even without applying any external strain, local strains caused by the atomic dislocation still occur in the vicinity of the GB. However, while the quantitative value of transmission across the boundary is slightly affected, its qualitative behavior around the gap as well as the transport gap is not strongly affected by this local strain. This can be explained by the fact that the transport gap is essentially due to the mismatch between electronic structures of graphene lattices in two sides (i.e., far from the grain boundary) where the effects of atomic dislocations is negligible.

In particular, the real space tight-binding Hamiltonian was constructed as follow: $H_{tb} = \sum_{i,j}t_{i,j} c_i^+ c_j$ where $t_{i,j}$ corresponds to the nearest-neighbor hoping energy between \textit{i}$^{th}$ and \textit{j}$^{th}$ atoms. When a uniaxial strain is applied, the change in the \textit{C-C} bond vector ${\vec r}_{ij}$ was determined according to ${\vec r_{ij}}\left( \vec \sigma\right) = \left\{ {\mathds{1} + {M_s}\left( {\vec \sigma} \right)} \right\}{\vec r_{ij}}\left( 0 \right)$ with the following strain tensor \cite{pere09}
\begin{equation}
{M_s}\left( {\vec \sigma} \right) = \sigma \left[ {\begin{array}{*{20}{c}}
	{{{\cos }^2}\theta  - \gamma {{\sin }^2}\theta }&{\left( {1 + \gamma } \right)\sin \theta \cos \theta }\\
	{\left( {1 + \gamma } \right)\sin \theta \cos \theta }&{{{\sin }^2}\theta  - \gamma {{\cos }^2}\theta }
	\end{array}} \right]
\end{equation}
where $\vec \sigma \equiv (\sigma, \theta)$ represents the strain (see in Fig.1) and $\gamma \simeq 0.165$ is the Poisson ratio \cite{blas70}. When taking into account strain effects, the hopping parameters were adjusted accordingly \cite{pere09}, i.e., $t_{ij}\left(\vec \sigma\right) = t_{ij}\left(0\right) \exp\left[ -3.37\left\{ r_{ij}\left(\vec \sigma\right)/a_0 - 1 \right\} \right]$ where $t_{ij}(0) \equiv - 2.7$ eV corresponds to the hoping energy between nearest neighbor atoms in pristine graphene. The tight-binding Hamiltonian $H_{tb}$ can be rewritten in its wavevector $k_y$-dependent (quasi-1D) form $H_{tb}(k_y)$, for more details see refs. \cite{chun14,hung15}. The Green's function was then computed \cite{anan08} using the equation: 
\begin{equation}
	\mathcal{G}(\epsilon, k_y) = [\epsilon  + i{0^+} - H_{tb}(k_y) - \Sigma(\epsilon, k_y)]^{-1},
\end{equation}
where the self-energy $\Sigma(\epsilon, k_y) = \Sigma_L(\epsilon, k_y) + \Sigma_R(\epsilon,k_y)$ with $\Sigma_{L,R}$ being the terms that describe the left and right contact-to-device couplings, respectively. The transmission probability needed to evaluate the current is calculated as $\mathcal{T}_e(\epsilon, k_y) = \rm{Tr}\left\{\Gamma_L\mathcal{G}\Gamma_R\mathcal{G}^{\dag}\right\}$, with $\Gamma_{L(R)} = i(\Sigma_{L(R)} - \Sigma_{L(R)}^{\dag})$ being the transfer rate at the left (right) contact. The conductance and current were obtained by the Landauer formulas:
\begin{eqnarray}
	G &=& \frac{e^2W}{\pi h}\int\limits_{BZ} dk_{y}\int d\epsilon \mathcal{T}_e(\epsilon,k_y)\left\lbrace -\frac{\partial f(\epsilon)}{\partial \epsilon} \right\rbrace , \\
	I &=& \frac{eW}{\pi h}\int\limits_{BZ} dk_{y}\int d\epsilon \mathcal{T}_e(\epsilon,k_y)\{f_L(\epsilon) - f_R(\epsilon)\}.
\end{eqnarray}
Here, the integrals over $k_y$ were performed in the whole Brillouin zone, $W$ denotes the channel width, and $f_{L(R)} = 1/[1+\exp((E-E_{FL(R)})/k_bT)]$ is the Fermi distribution function in the left (right) contact with the Fermi energy level $E_{FL(R)}$.
\begin{figure*}[!b]
	\centering
	\includegraphics[width = 0.9\textwidth]{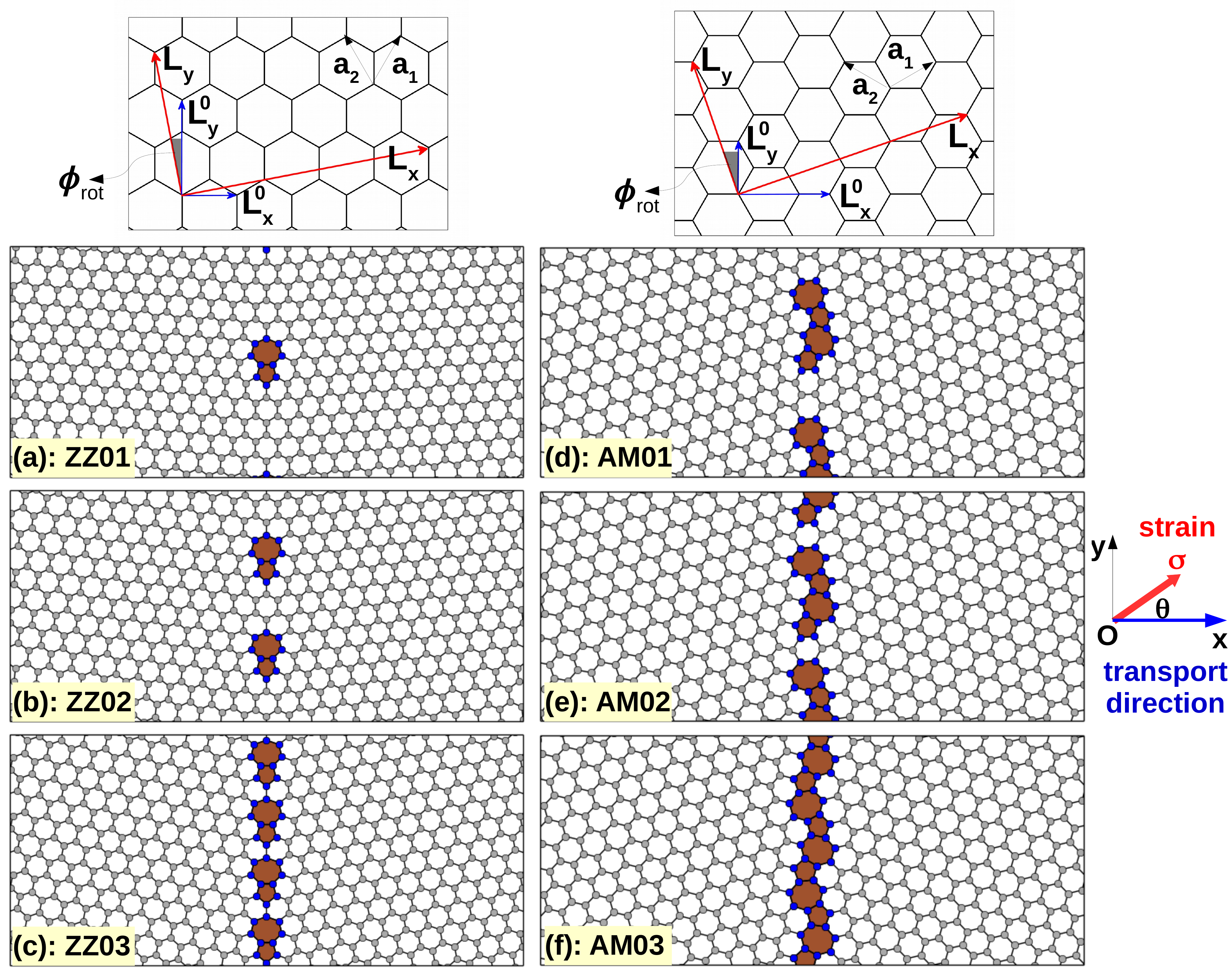}
	\caption{Graphene-based systems investigated in this work. The systems contain a single grain boundary that presents periodic pentagon-heptagon pairs along the Oy axis. The top imagines illustrate the rotation of graphene sheet to form two graphene domains surrounding the grain boundary (see text). The transport direction (Ox axis) is perpendicular to the grain boundary. The uniaxial strain of magnitude $\sigma$ is applied in the direction $\theta$ with respect to the Ox axis as illustrated by the small inset on the right.}
	\label{fig_sim0}
\end{figure*}

\section{Results and discussion}

Within the methodology described above, we investigate the transport properties through different graphene grain boundary systems. 

\subsection{Transport gap induced by strain}

First, we investigate the transport properties of the symmetrical graphene-based systems schematized in Fig.1. Considering a strain applied along the Ox axis, the symmetrical graphene GBs have been found to always remain semi-metallic \cite{kuma12}. However, this conclusion is no more true when the strain direction is changed. The strain effects on the transmission function obtained in the system with the ZZ$_{21}$ and ZZ$_{12}$ lattices surrounding the GB (see Fig.1.c) are represented in Fig.2. It is true that in the unstrained case and for a strain $(\sigma,\theta) = (4\%,0^\circ)$, the system remains semi-metallic with a zero enegy gap in its transmission function. However, when the strain direction is changed, a significant gap $E_g$ opens, as shown in Figs.2.c-d. The origin of this feature can be easily explained as follows. In principle, the energy gap in the transmission function is proportional to the separation between the Dirac cones of the two graphene domains $\Delta D_y = |D_{1y} - D_{2y}|$ along the $k_y$-axis \cite{chun14,hung15}. Since the two graphene domains are commensurable in this case, their electronic structures are similar, i.e., their Dirac cones are located in the same positions along the $k_y$-axis if no strain is applied. This explains the zero transport gap observed in the unstrained system. When a strain is applied, the position of Dirac cones generally changes. However, since the system is symmetric with respect to the grain boundary, $\Delta D_y = 0$ and hence the gap remains zero if the strain is applied along the Ox direction ($\theta = 0^\circ$). For other directions, in particular $\theta = 20^\circ$ and $45^\circ$  as shown in Figs.2.c-d, the symmetry property of the system is broken by strain and hence a finite $\Delta D_y$ and $E_g$ are achieved.
\begin{figure*}[!b]
	\centering
	\includegraphics[width = 0.82\textwidth]{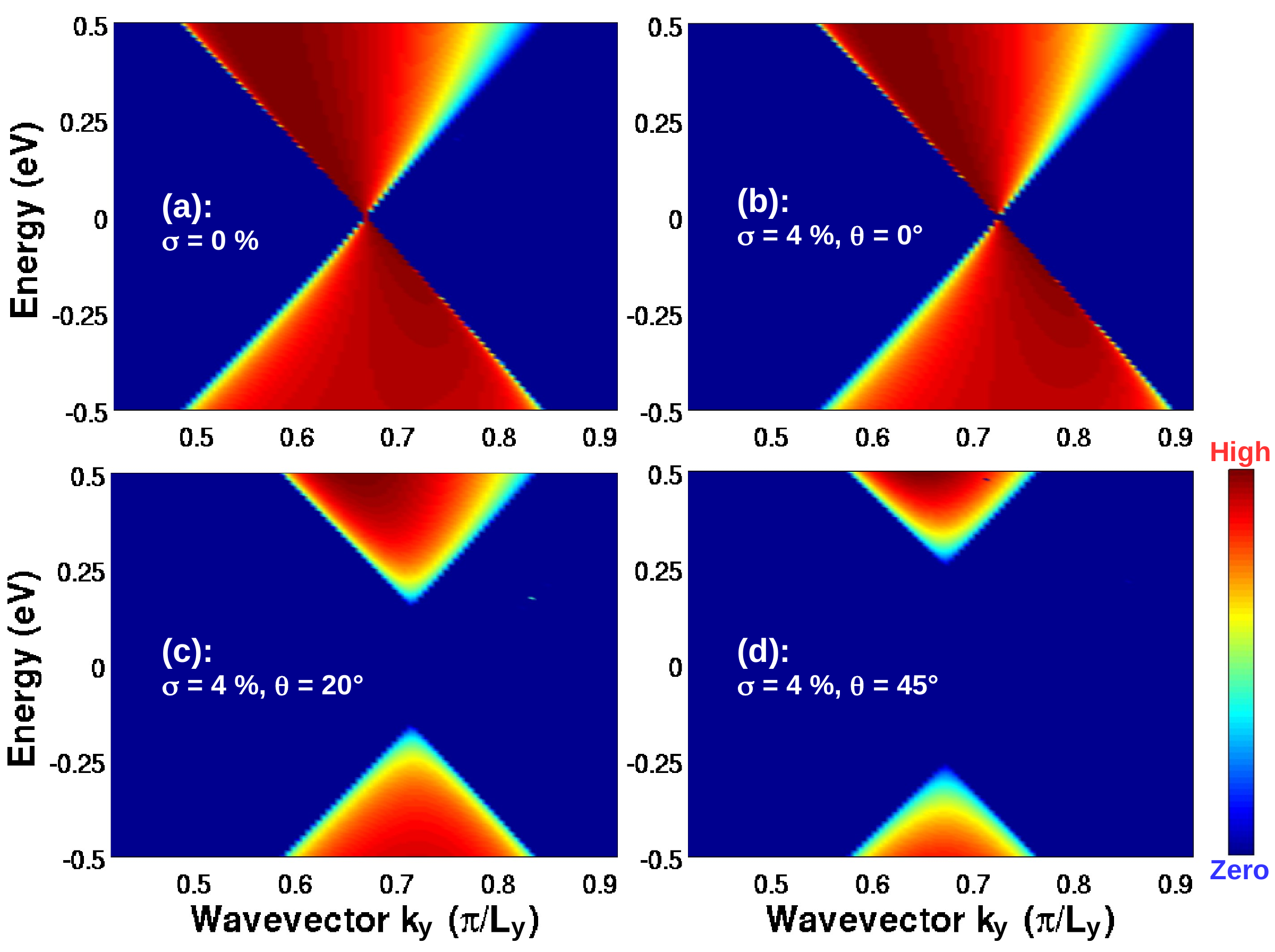}
	\caption{Transmission probability maps around the neutrality (Dirac) point obtained in the graphene GB system ZZ03 (see Fig.1.c). Various applied strains ($\sigma,\theta$) are considered.}
	\label{fig_sim1}
\end{figure*}

In Fig.3, maps of transport gap with respect to the strain magnitude (from $0\%$ to 6$\%$) and its applied direction were calculated for the different graphene systems described in Fig.1. Except for the AM02 case that will be discussed later, the transport gap is zero for $\theta = i \times 90^\circ$ (where $i$ = 0,1,2,3) while it gets maximums around $\theta = i \times 90^\circ + 45^\circ$. In conclusion, opening a finite transport gap in graphene GB systems is always possible even if they are symmetric. The gap value is strongly dependent on both the strain magnitude and its applied direction. More interestingly, when strain is applied in the appropriate direction, a large transport gap of a few hundred meV can be achieved with strain as small as a few percents, e.g., a large gap up to 840 meV is achieved for $\sigma = 6 \%$ as shown in Fig.3.b. 
\begin{figure*}[!b]
	\centering
	\includegraphics[width = 0.99\textwidth]{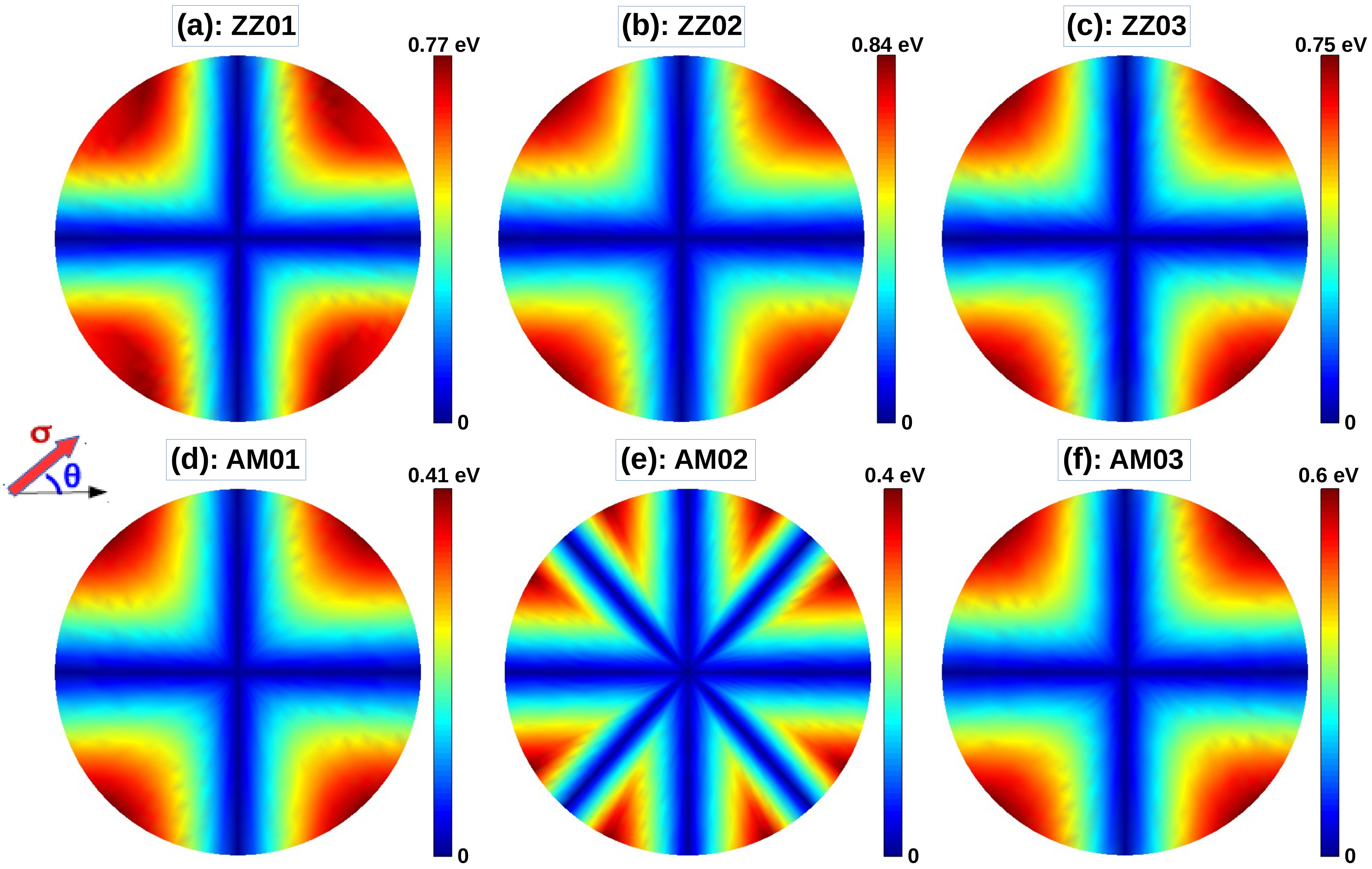}
	\caption{Transport-gap maps with respect to the strain $(\sigma,\theta)$ in different graphene GB systems (see Fig.1). Note that the radius from the central point of the maps represents the strain magnitude ranging from 0$\%$ (center) to 6$\%$ (edge).}
	\label{fig_sim2}
\end{figure*}

\begin{figure*}[!b]
	\centering
	\includegraphics[width = 0.78\textwidth]{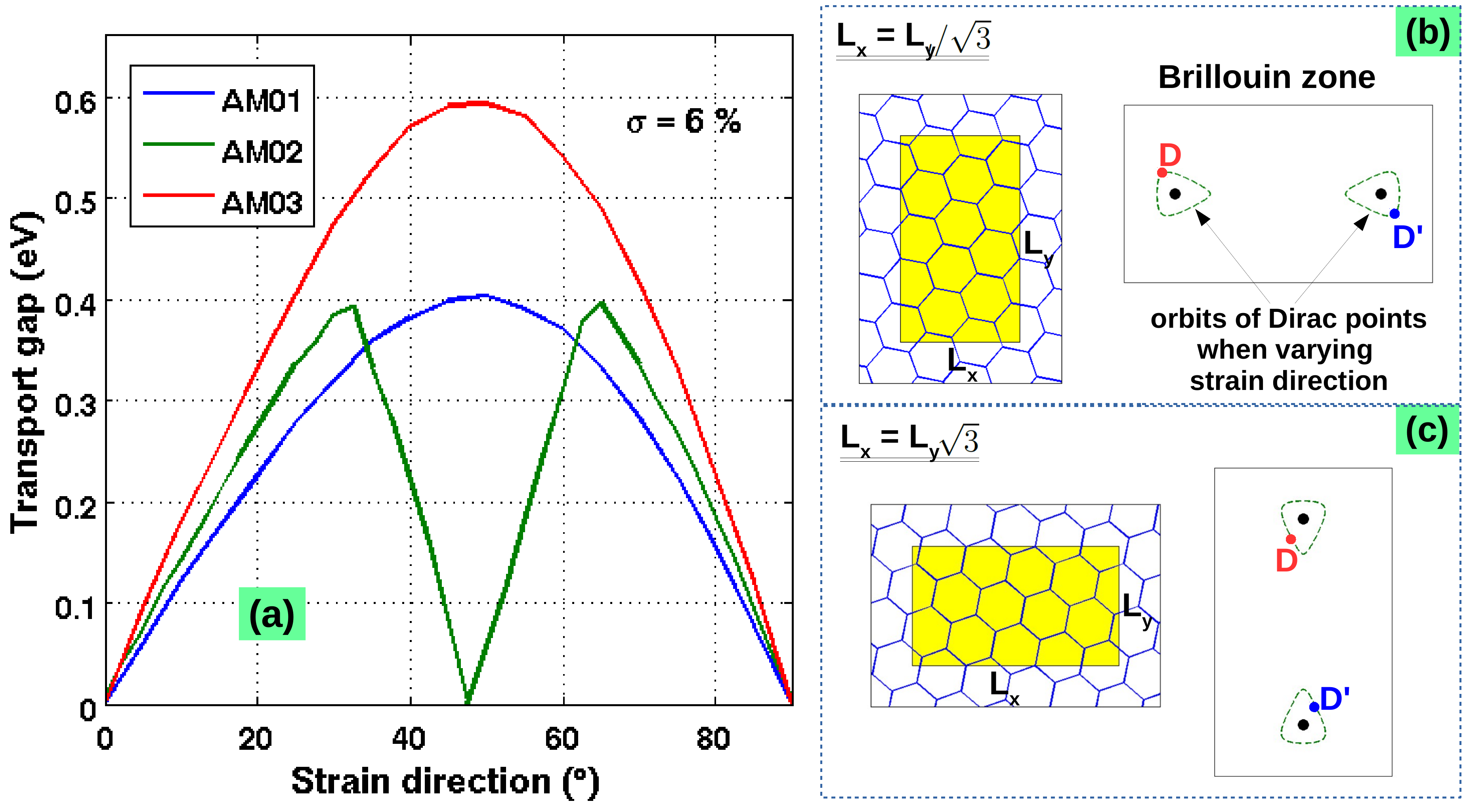}
	\caption{Transport gap versus strain direction $\theta$ in armchair oriented graphene based systems for $\sigma = 6\%$ (a). Supercells and Brillouin zone for the two typical graphene lattices are illustrated in (b) and (c).}
	\label{fig_sim3}
\end{figure*}

Moreover, Fig.3 also demonstrates that besides its dependence on the strain magnitude and strain direction, the transport gap is also very sensitive to the lattice symmetry. First, the gap is shown to be significantly dependent on the misoriented angle $\phi_{MO}$ between two graphene domains. Indeed, for zigzag-oriented lattices and for small strains, the gap generally decreases when increasing the misoriented angle as shown for instance in the cases ZZ02 ($\phi_{MO}\approx13.2^\circ$) - Fig.3.b and ZZ03 ($\phi_{MO}\approx21.8^\circ$) - Fig.3.c. However, for armchair-oriented lattices, the gap has an opposite behavior, which is seen in the cases AM01 ($\phi_{MO}\approx17.9^\circ$) - Fig.3.d and AM03 ($\phi_{MO}\approx27.8^\circ$) - Fig.3.f. Here, we would like to notice that the bandstructure of graphene domains surrounding the grain boundary is actually obtained as the folding bands from the original lattice. In principle, the band folding is essentially dependent on both the rotation of graphene sheet and its original lattice. In the original lattices, the Dirac cones are located at different positions in the $k-$space, i.e., at $(\mp \frac{2\pi}{3L_x}, 0)$ and $(0,\mp \frac{2\pi}{3L_y})$ in the zigzag- and armchair oriented cases, respectively. Hence, the folding bands of corresponding rotated lattices have different behaviors in these two cases when strain is applied, which can explain the results discussed above.

Next, the behavior of transport gap is also found to be strongly dependent on other lattice symmetry properties, i.e., the supercell (shape and size) of graphene domains. In Fig.4.a, the transport gap is presented as a function of strain direction $\theta$ obtained for $\sigma = 6\%$ in the armchair-oriented cases. Actually, the transport gap in the AM02 system ($\phi_{MO} \approx 21.8^\circ$) obtained for strain directions close to $\theta = i \times 90^\circ$ also satisfies the $\phi_{MO}$-dependence discussed above for the armchair-oriented systems. However, in contrast with the other cases, it exhibits an additional valley, instead of a maximum, around $\theta = i \times 90^\circ + 45^\circ$ (see also Fig.3.e). To explain this feature, the supercell and Brillouin zone of two typical graphene lattices are illustrated in Figs.4.b-c. The displacement of the Dirac cones when changing the strain direction is also depicted. In principle, the supercell of graphene lattices has two possible shapes, i.e., either $L_x = L_y/\sqrt{3}$ or $L_x = L_y\sqrt{3}$. This property is essentially originated from the relationship between $\mathbf{L}_x$ and $\mathbf{L}_y$ described in the previous section. Accordingly, there are two different shapes of its Brillouin zone and the corresponding \textit{K}-points (or Dirac points without strain) are located either at $(\mp \frac{2\pi}{3L_x}, 0)$ or at $(0,\mp \frac{2\pi}{3L_y})$, respectively (see Figs.4.b-c). In the AM01, AM03, ZZ01, ZZ02 and ZZ03 - systems (i.e., $L_x = L_y\sqrt{3}$), the transport gap $E_g$ is simply proportional to $\Delta D_y = \left| D_{1y} - D_{2y}\right|$ ($\equiv \left| D_{1y}' - D_{2y}'\right|$). Hence, $E_g$ presents maximums around $\theta \approx i \times 90^\circ + 45^\circ$ while it is zero for $\theta = i \times 90^\circ$ when $D_{1y} \equiv D_{2y}$ and $D_{1y}' \equiv D_{2y}'$. However, the situation is different for the AM02 systems (i.e., $L_x = L_y/\sqrt{3}$) since $\Delta D_y$ is the minimum of $\left| D_{1y} - D_{2y}\right|$, $\left| D_{1y} - D_{2y}'\right|$, $\left| D_{1y}' - D_{2y}\right|$ and $\left| D_{1y}' - D_{2y}'\right|$, making twice smaller the cycle of $\theta$-values for which the transport gap vanishes. Hence, besides the zero value of $E_g$ at $\theta = i\times90^\circ$, the gap is also zero with a corresponding valley around additional points $\theta \approx i\times90^\circ + 45^\circ$ when  $\left| D_{1y} - D_{2y}'\right|$ ($\equiv \left| D_{1y}' - D_{2y}\right|$) is zero or small. Note that this dependence of $E_g$ on the supercell shape is a common property for all graphene GB systems even if (not shown in this paper) graphene domains are asymmetric (but still commensurable) or incommensurable.

In order to investigate the dependence on the supercell size, the transport gap is plotted as a function of strain direction $\theta$ in Fig.5.a and strain magnitude in Fig.5.b obtained in the case of zigzag-oriented systems. Note that the transport gap in the ZZ01 system ($\phi_{MO} \approx 9.4^\circ$) obtained for small strains or large strains with directions close to $\theta = i\times 90^\circ$ also satisfies the $\phi_{MO}$-dependence discussed above for the zigzag-oriented systems (see Fig.5.a). However, when the strain magnitude is large enough, the gap around $\theta = 45^\circ$ obtained in this system is suddenly reduced when increasing $\sigma$. To clarify this feature, the transport gap is investigated in a large range of $\sigma$ while $\theta$ is fixed at $45^\circ$ (Fig. 5.b). Actually, in all cases, the gap is found to increase with the strain magnitude and then to decrease when $\sigma$ is large enough. The gap can even vanish at $\sigma \simeq 14.3\%$ in the case of ZZ01 system (Fig.5.b). Interestingly, the threshold value of strain $\sigma_{peak}$ at which the behavior of $E_g$ changes is found to be inversely proportional to the size $L_y$ of the supercell. In particular, $\sigma_{peak}$ is $\sim$ 11.2 $\%$, 6.7 $\%$ and 4.9 $\%$ for $L_y \simeq 0.65$ nm (ZZ03), $1.1$ nm (ZZ02) and $1.5$ nm (ZZ01), respectively. To explain these properties, diagrams to illustrate visually the displacement of the Dirac cones of two graphene domains along the $k_y$ axis when a large strain is applied are drawn in Fig.5.c. For small strains, the Dirac cones $D_1$ (of domain 1) and $D_2$ (of domain 2) are gradually separated inducing the transport gap to increase with the strain magnitude. When the strain is large enough, $D_1$ reaches the edge of Brillouin zone and then the situation (1) illustrated in Fig.5.c takes place, thus modifying the behavior of $\Delta D_y$ (and $E_g$) from increasing to decreasing with $\sigma$. With further increasing strain, $D_2$ reaches the zero point inducing the situation (2) to happen which makes the gap decreasing even more rapidly. Additionally, the dependence of $\sigma_{peak}$ on $L_y$ is a direct consequence of the $L_y$-dependence of the size of the Brillouin zone. For smaller $L_y$, the corresponding Brillouin zone is larger and hence the situation described above occurs at larger strain. 
\begin{figure*}[!b]
	\centering
	\includegraphics[width = 0.8\textwidth]{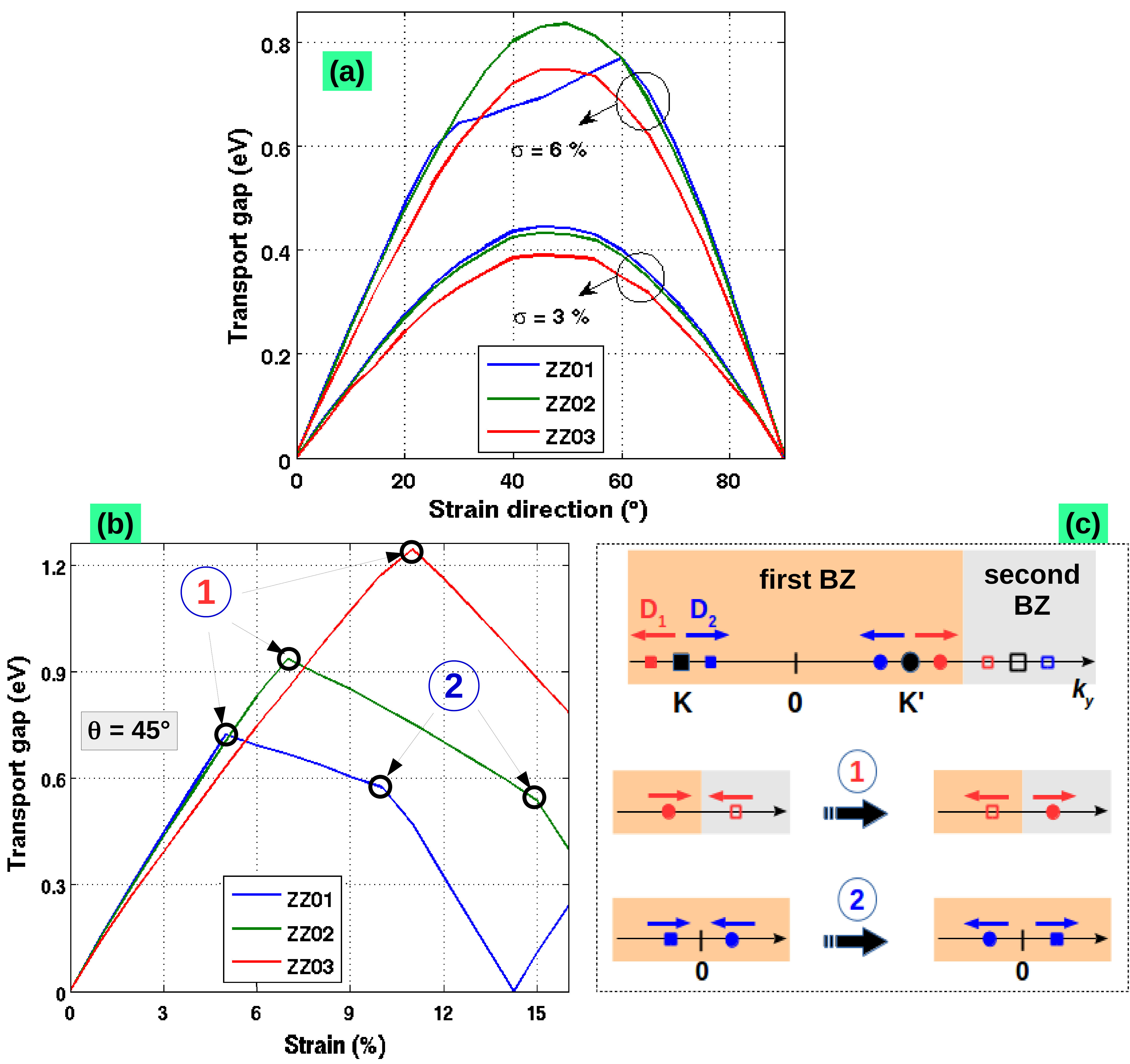}
	\caption{Transport gap versus strain direction $\theta$ for two different $\sigma$ values (a) and versus $\sigma$ for $\theta = 45^\circ$ (b) for zigzag oriented graphene systems. (c) Schematics illustrating the displacement induced by srain along the $k_y$ axis of the Dirac points related to the two graphene domains (see text).}
	\label{fig_sim3}
\end{figure*}

We would like to notice that such a merging of Dirac cones when a large uniaxial strain is applied in pristine graphene has already been reported in \cite{pere09} showing that beyond the critical strain ($\sim 23\%$), a bandgap opens. However, the merging process in the present work is a completely different process. Indeed, it appears at much lower strain than the critical value mentioned above and the bandstructure of graphene lattices is still gapless. This is because the presence of grain boundary changes the periodicity of graphene lattices and the merging process observed here is essentially a consequence of the folding of the graphene bandstructure from the original Brillouin zone to the reduced Brillouin one. Consequently, our predictions suggest that the graphene GB systems with a short periodic length $L_y$ are preferable to obtain an increase of transport gap in a large range of strain.
\begin{figure*}[!t]
	\centering
	\includegraphics[width = 0.8\textwidth]{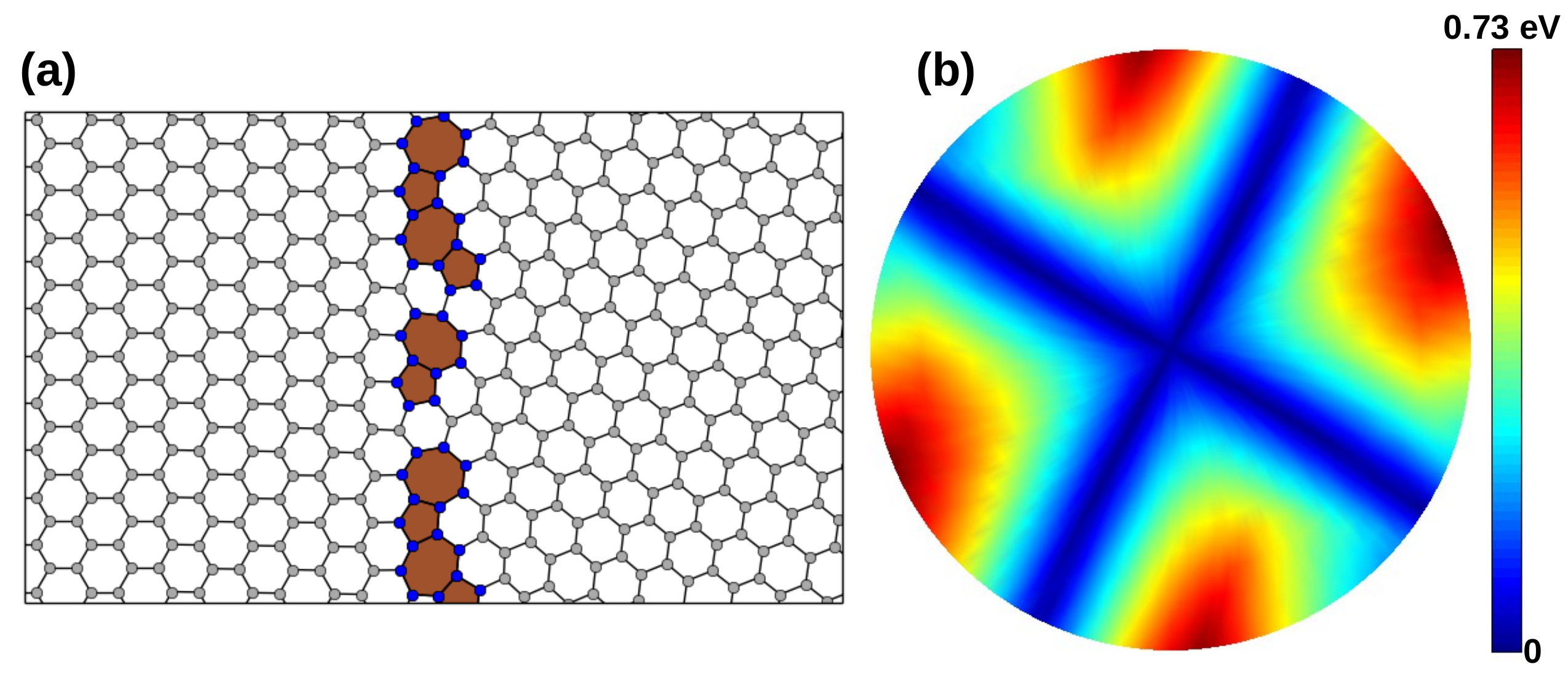}
	\caption{(a) Asymmetrical graphene GB system: an armchair oriented lattice (on the left) connected to a ZZ$_{35}$ lattice (on the right). (b) Map of transport gap with respect to the strain ($\sigma,\theta$). The radius from the central point represents the strain magnitude $\sigma$ ranging from 0$\%$ (center) to 6$\%$ (edge).}
	\label{fig_sim3}
\end{figure*}

In order to present a general approach, the properties of $E_g$ in asymmetrical (but still commensurable) systems are also investigated. Since the gap is dependent on both the strain and orientation of graphene domains, the properties of $E_g$ should change dramatically when the global system is no longer symmetric. In Fig.6.b, a map of $E_g$ with respect to the strain was calculated for the asymmetrical system represented in Fig.6.a. Note that $E_g$ has properties similar to that of symmetrical systems, except that it exhibits different $\theta$-dependence. In particular, the gap is strongly dependent on the strain direction with valleys of minimal values at $\theta \simeq i \times 90^\circ + 61^\circ$ and high peaks of maximal values in between such valleys. A large gap up to 730 meV for strain $\simeq 6\%$ is also observed. 
\begin{figure*}[!t]
	\centering
	\includegraphics[width = 0.99\textwidth]{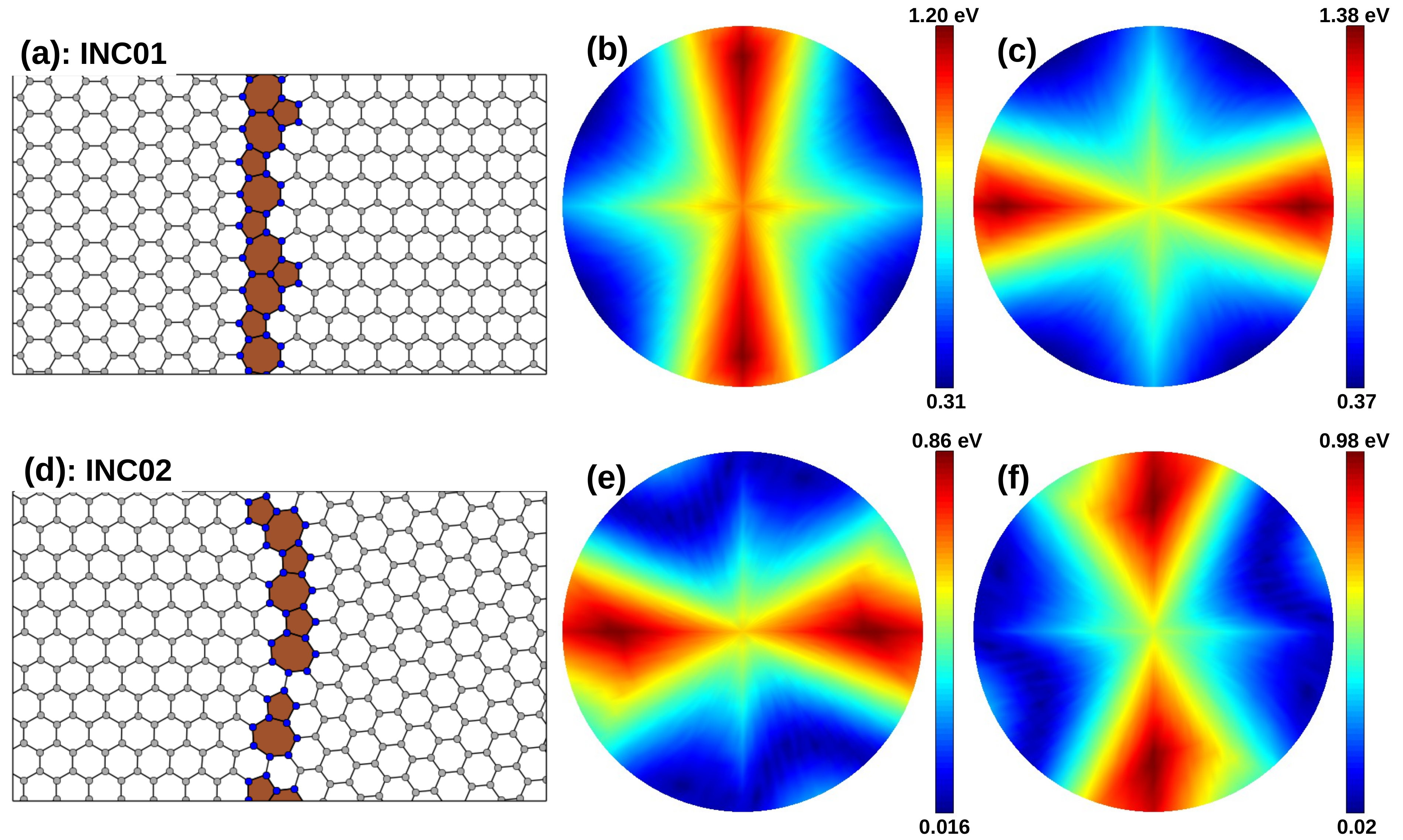}
	\caption{Maps (b,c) and (e,f) of transport gap with respect to the strain ($\sigma,\theta$) in two incommensurable systems shown in (a) and (d), respectively. Both tensile and compressive strains are considered in (a,c) and (b,d), respectively. The radius from the central point represents the strain magnitude $\sigma$ ranging from 0$\%$ (center) to 6$\%$ (edge).}
	\label{fig_sim3}
\end{figure*}

The next section will be devoted to the effects of strain on the transport gap in incommensurable graphene-based systems. Note that in the incommensurable systems where the two graphene domains have the same supercell shape (not shown in this paper), the transport gap has properties similar to that of commensurable ones discussed above. However, differently from the commensurable systems, it is possible to design heterostructures where the graphene domains have different supercell shapes (see Figs.4.b-c) in the incommensurable cases. In these heterostructures, the Dirac cones of the two graphene domains are located at different positions in the $k$-space (i.e., at $(\mp \frac{2\pi}{3L_x}, 0)$ and $(0,\mp \frac{2\pi}{3L_y})$). Consequently, a finite transport gap can be achieved even if no strain is applied \cite{yazy10}. In Fig.7, the maps of $E_g$ with respect to the strain are displayed for two different incommensurable systems shown in Fig.7.a (INC01) and Fig.7.d (INC02), respectively. Here, both the tensile (Fig.7.b and Fig.7.e) and compressive strains (Fig.7.c and Fig.7.f) are investigated. When no strain is applied, the gaps of $\sim$ 1 eV and 0.58 eV are obtained for the INC01 and INC02 systems, respectively. More interestingly, strain engineering is found to be a very efficient way to modulate the gap. Depending on the strain direction, gaps in the range of $\left[ 0.31, 1.20\right]$ eV and $\left[ 0.37, 1.38\right]$ eV can be respectively achieved when the tensile and compressive strains ($\leq 6\%$) are applied to the INC01 system. In the case of INC02, the ranges of $E_g$ are $\left[ 0.016, 0.86\right]$ eV (tensile strain) and $\left[ 0.02, 0.98\right]$ (compressive strain) eV. Additionally, the behavior of $E_g$ is similar in two strain cases ($\sigma,\theta$) and ($-\sigma,\theta+90^\circ$). This can be explained by analyzing the strain tensor in Eq. (1). For small strains, the relationship between the bond lengths under these two strains is approximately given by 
\begin{equation}
	r(\sigma,\theta) - r(-\sigma,\theta + 90^\circ) \simeq \sigma (1 - \gamma) r_0, \nonumber
\end{equation}
which is $\theta$-independent for all \textit{C-C} bond vectors. It also implies that there is a fixed ratio between the hopping energies
$t_{ij}(\sigma,\theta)$ and $t_{ij}(-\sigma,\theta + 90^\circ)$ and hence the displacement of Dirac cones (and $E_g$) are qualitatively similar in these two cases \cite{chun14}. This property is actually valid for all the cases, i.e., including also the commensurable systems studied above. The only difference between the effects of the two types of strain is that the gap varies more strongly with the compressive strain than with the tensile one.

Finally, we would like to emphasize that the strain-induced finite transport gap only occurs in systems where two graphene domains have different orientations. In systems presenting the same oriented graphene lattices (e.g., graphene sheet with a line defect \cite{chen14}), the gap is always be zero because the strain effects on their electronic structure are the same and hence can not lead to the separation of their Dirac cones in the \textit{k}-space. 

\subsection{Strain versus defect scattering}

In polycrystalline graphene, the scattering on the GB defects is one of important factors that strongly affects its transport properties (carrier mobility, mean free path, conductivity ... \cite{song12,tuan13,jime14}). In this subsection, we investigate these scattering mechanisms when strain is applied. Our calculations have shown that the conductance of graphene is drastically affected and strongly degraded in the presence of these GB defects. However, in the systems where the two graphene domains have a supercell with the shape as shown in Fig.4.b, e.g., the AM02 system (Fig.1.b), the strain effects can significantly modify the GB defect scattering. In Fig.8.a, the conductance as a function of Fermi energy was calculated in the AM02 system for different applied strains and compared to the unstrained case. To avoid finite transport gap effects, only strains of $\theta = 0^\circ$ and $90^\circ$ are considered here. The conductance around the neutrality point is shown to be significantly enhanced when a strain is applied. Indeed, its value is about twice the conductance obtained in the unstrained case. When a strain is applied, the bandstructure of both graphene domains is modified and their band profiles around the \textit{K} and \textit{K'} points are separated along the $k_y$ axis as illustrated in Fig.8.b. Interestingly, the enhancement of conductance is especially significant in the energy window $\left[E_1,E_2\right]$  where there is no overlap between the bands around \textit{K} and \textit{K'}. Beyond this energy window, the conductance enhancement weakens gradually when increasing the energy.
\begin{figure*}[!t]
	\centering
	\includegraphics[width = 0.85\textwidth]{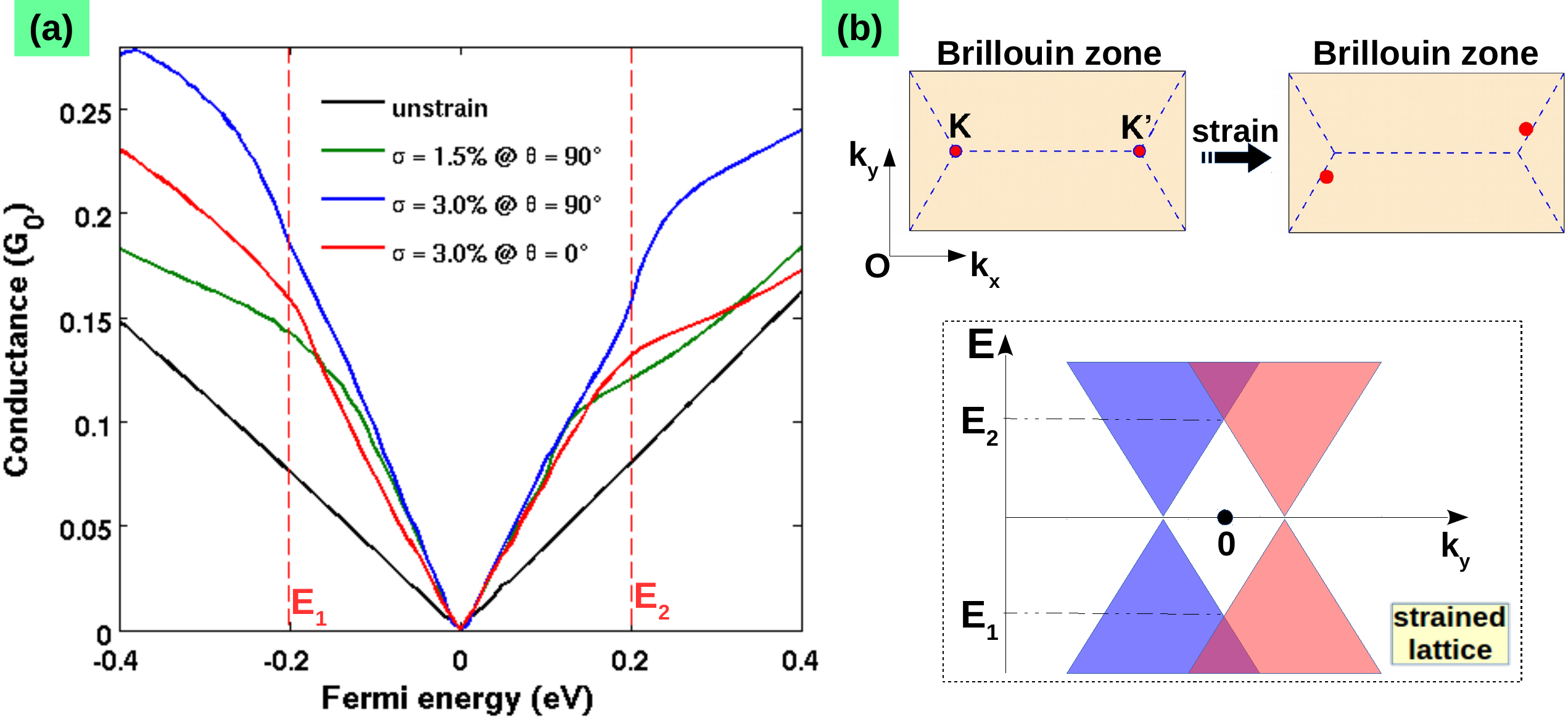}
	\caption{Zero-temperature conductance as a function of Fermi energy in the AM02 system (see Fig.1.e) for different applied strains (a). $G_0 = e^2W/hL_y$. (b) Schematics illustrating the Brillouin zone and band profiles along the $k_y$ axis for two graphene domains along the $k_y$-axis when strain is applied.}
	\label{fig_sim3}
\end{figure*}

In order to explain this peculiar phenomenon, schematic diagrams illustrating the transmission/reflection processes, band profiles of both graphene domains, and phase difference $\Delta\phi$ between the incoming and reflected states computed at the atomic sites are presented in Fig.9. When transmitting through the graphene system, the incoming wave $e^{ikx}$ will always be separated at the grain boundary into two components $te^{ikx}$ and $re^{-ikx}$ that correspond to the transmitted and reflected waves, respectively. The transmission and reflection probabilities of this process are simply determined by $T = \left| t \right|^2$ and $R = \left| r \right|^2$. If there is no scattering source (i.e., no defect), $T = 1$ and $R = 0$. The presence of defects in the GB should enhance the transition between the incoming and reflected states (i.e., increase of the amplitude of $r$) and hence enlarge the reflection. The transition probability of these processes is, in principle, dependent on both the nature of the defects and the properties of these two states.

Here, we would like to remind a peculiar phenomenon in graphene: the parity symmetry of the subbands in zigzag graphene nanoribbons (ZGNRs) \cite{cres08}. Because of their inversion symmetry, the subbands of ZGNRs have alternatively even/odd parity wavefunctions. These properties also imply that $\left| \left\langle \psi^K | \psi^{K'} \right\rangle \right| \simeq 0$ if these states have the same parity symmetry while $\left| \left\langle \psi^K | \psi^{K'} \right\rangle \right|$ is high when they have different parities. In ZGNR p--n junctions, the superimposed potential acts as an intervalley scattering source. However, because of the parity symmetries only the transition between the states of the same parity is allowed while it is blocked for the states of different parities. This is the so-called parity selective tunneling in ZGNR devices. Interestingly, similar properties are found in the system studied here. In particular, as illustrated in Fig.9, while the transition processes from the $\left| p^+ \right\rangle$ states in the left domain to the states $\left| q^+ \right\rangle$ in the right one contribute to the transmission, the processes from $\left| p^+ \right\rangle$ to $\left| q^- \right\rangle$ correspond to the reflection. Fig.9.c shows that while the states $\left| p_1^+ \right\rangle$ and $\left| q_2^- \right\rangle$ (similarly, between $\left| p_2^+ \right\rangle$ and $\left| q_1^- \right\rangle$) have the same symmetry property (i.e., with the phase difference $\Delta\phi \simeq 0$), $\Delta\phi$ between the states $\left| p_1^+ \right\rangle$ and $\left| q_1^- \right\rangle$ (similarly, between $\left| p_2^+ \right\rangle$ and $\left| q_2^- \right\rangle$) varies strongly along the atomic position. This leads to the fact that their projection is high in the former case while it is very small in the latter one, similarly to the properties of the even/odd parity subbands in ZGNRs as mentioned above. These properties suggest that the strain-induced degradation of the defect scattering (i.e., the conductance enhancement) can be understood as follows. First, the defect scattering tends to promote the transition between the states $\left| p^+ \right\rangle$ and $\left| q^- \right\rangle$ and hence to reduce significantly the conductance of the GB systems. In the unstrained case, all four processes $\left| p_{1}^+ \right\rangle \rightarrow \left| q_{1,2}^- \right\rangle$ and $\left| p_{2}^+ \right\rangle \rightarrow \left| q_{1,2}^- \right\rangle$ contribute to the reflection. When strain is applied, the Dirac cones $D$ and $D'$ are separated along the $k_y$ axis and hence the processes $\left| p_{1}^+ \right\rangle \rightarrow \left| q_{2}^- \right\rangle$ and $\left| p_{2}^+ \right\rangle \rightarrow \left| q_{1}^- \right\rangle$ are forbidden. Due to their symmetry properties, the transition probability of the processes $\left| p_{1}^+ \right\rangle \rightarrow \left| q_{2}^- \right\rangle$ and $\left| p_{2}^+ \right\rangle \rightarrow \left| q_{1}^- \right\rangle$ (i.e., between states of the same symmetry) should be much larger than that for $\left| p_{1}^+ \right\rangle \rightarrow \left| q_{1}^- \right\rangle$ and $\left| p_{2}^+ \right\rangle \rightarrow \left| q_{2}^- \right\rangle$ (i.e., between states of different symmetries). Consequently, the disappearance of the former processes is suggested to be the principal reason of the degradation of GB defect scattering (i.e., the conductance enhancement) in the strained system, compared to the unstrained case. This also explains why the conductance enhancement gradually weakens at higher energies beyond the energy window $\left[E_1,E_2\right]$, where the overlap between the $D$ and $D'$ cones occurs again as in the unstrained system. Note that because of their lattice symmetry, the properties discussed above do not occur in systems where the supercells of graphene domains are as shown in Fig.4.c.
\begin{figure*}[!t]
	\centering
	\includegraphics[width = 0.88\textwidth]{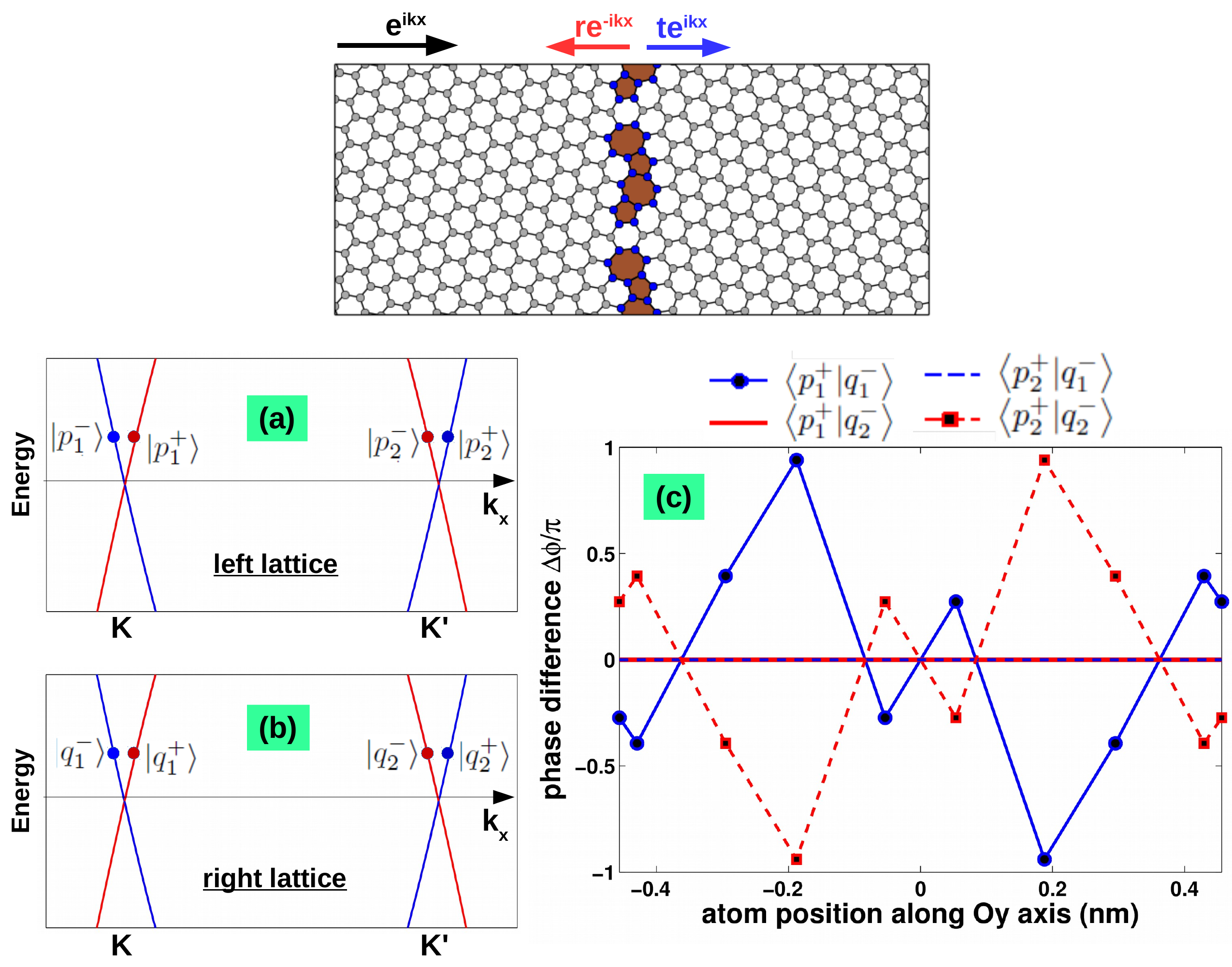}
	\caption{The top panel illustrates transmission and reflection processes at the grain boundary. Two diagrams (a,b) depict the band profile of graphene domains along the $k_x$-axis. The phase difference $\Delta \phi$ between the incoming state $\left| p^+ \right\rangle$ and reflected state $\left| q^- \right\rangle$ along the Oy axis is displayed in (c).}
	\label{fig_sim3}
\end{figure*}

Together with our previous findings on the strain-induced transport gap, the results obtained here demonstrate the decisive role of the lattice symmetry in the transport picture of graphene devices, thus motivating further investigations on these graphene hetero-channels in view of using strain engineering to restrain the detrimental impact of defects on the transport properties of polycrystalline graphene devices \cite{song12,tuan13,jime14}.

\subsection{Suggested applications}

In the previous sections, we demonstrated that (i) the transport properties of graphene GB systems are very sensitive to the strain (both in magnitude and direction), (ii) strain engineering can be used to open a finite transport gap in all systems where the two graphene domains have different orientations and (iii) a large gap of about a few hundred meV can be achieved with a strain as small as a few percents. Unquestionably, thanks to such the properties, these graphene heterostructures should be very interesting for practical applications.
\begin{figure*}[!b]
	\centering
	\includegraphics[width = 0.6\textwidth]{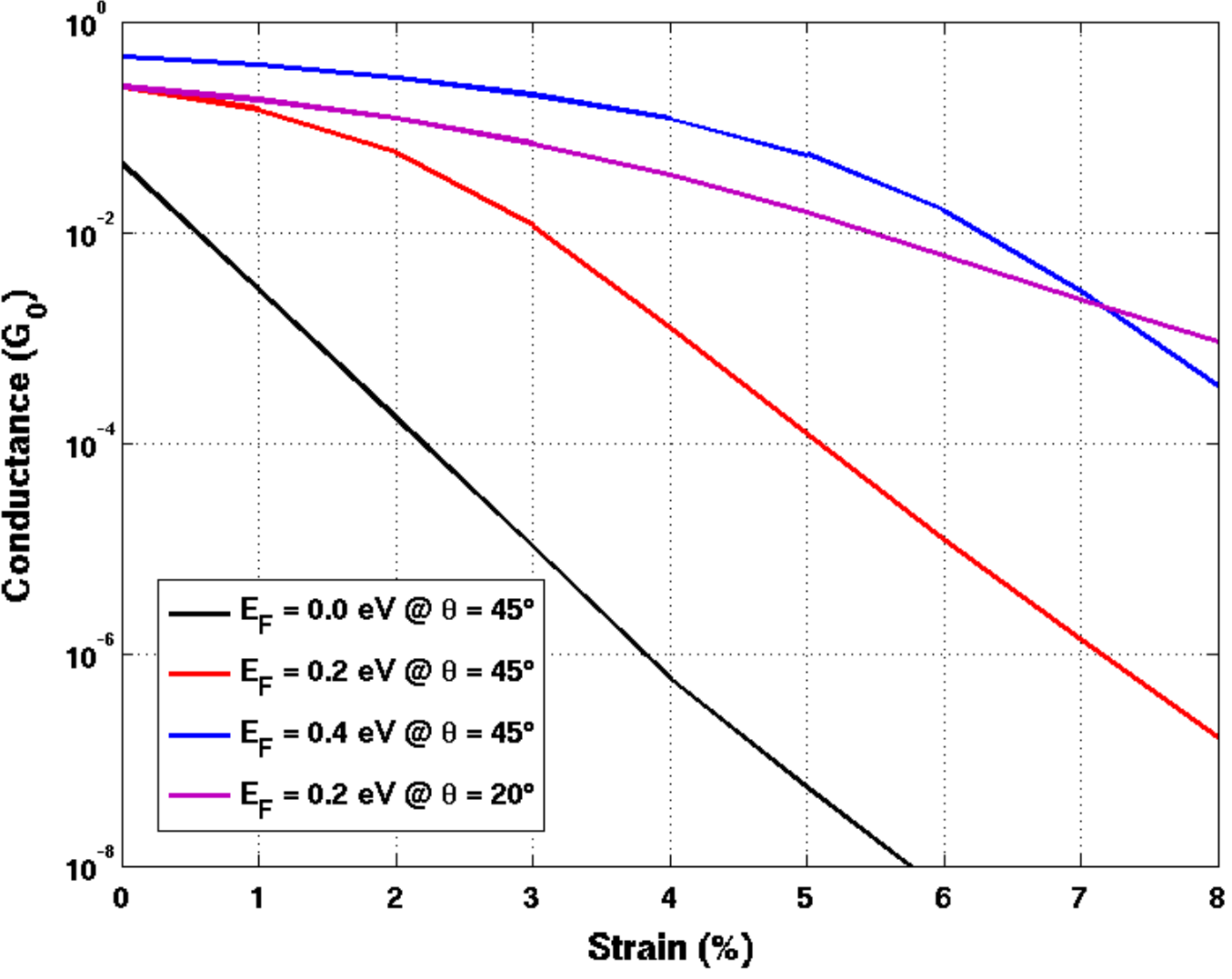}
	\caption{Conductance in the ZZ03 system (see Fig.1.c) as a function of strain magnitude for different Fermi levels and strain directions. $G_0 = e^2W/hL_y$. The simulations were performed at room temperature.}
	\label{fig_sim3}
\end{figure*}

Room-temperature conductance of the ZZ03 system (see in Fig.1) as a function of strain magnitude obtained for different Fermi levels and strain direction is presented in Fig.10. Indeed, due to its strong sensitivity to the strain, the conductance can be largely modulated when varying the strain magnitude (range of a few percents) and strain direction. In particular, the conductance can be reduced by a factor of $\sim 10^5$ and $10^3$ at $E_F = 0$ and $0.2$ eV, respectively, for a small strain of $\sim 4 - 5\%$. The conductance can also be significantly modulated when changing the strain direction from $\theta = 20^\circ$ to $45^\circ$ for $E_F = 0.2$ eV. This result suggests the possible use of this graphene system in highly sensitive strain nano-sensors. Given the large strain-induced gap, the system could be also very promising for the applications in flexible transistors. Indeed, a room-temperature ON/OFF ratio (i.e., $G_{E_F = 0.4 eV}/G_{E_F = 0}$) as large as a few tens of thousands can be achieved when a strain of only $3 - 4\%$ is applied, as shown in Fig.10.
\begin{figure*}[!b]
	\centering
	\includegraphics[width = 0.9\textwidth]{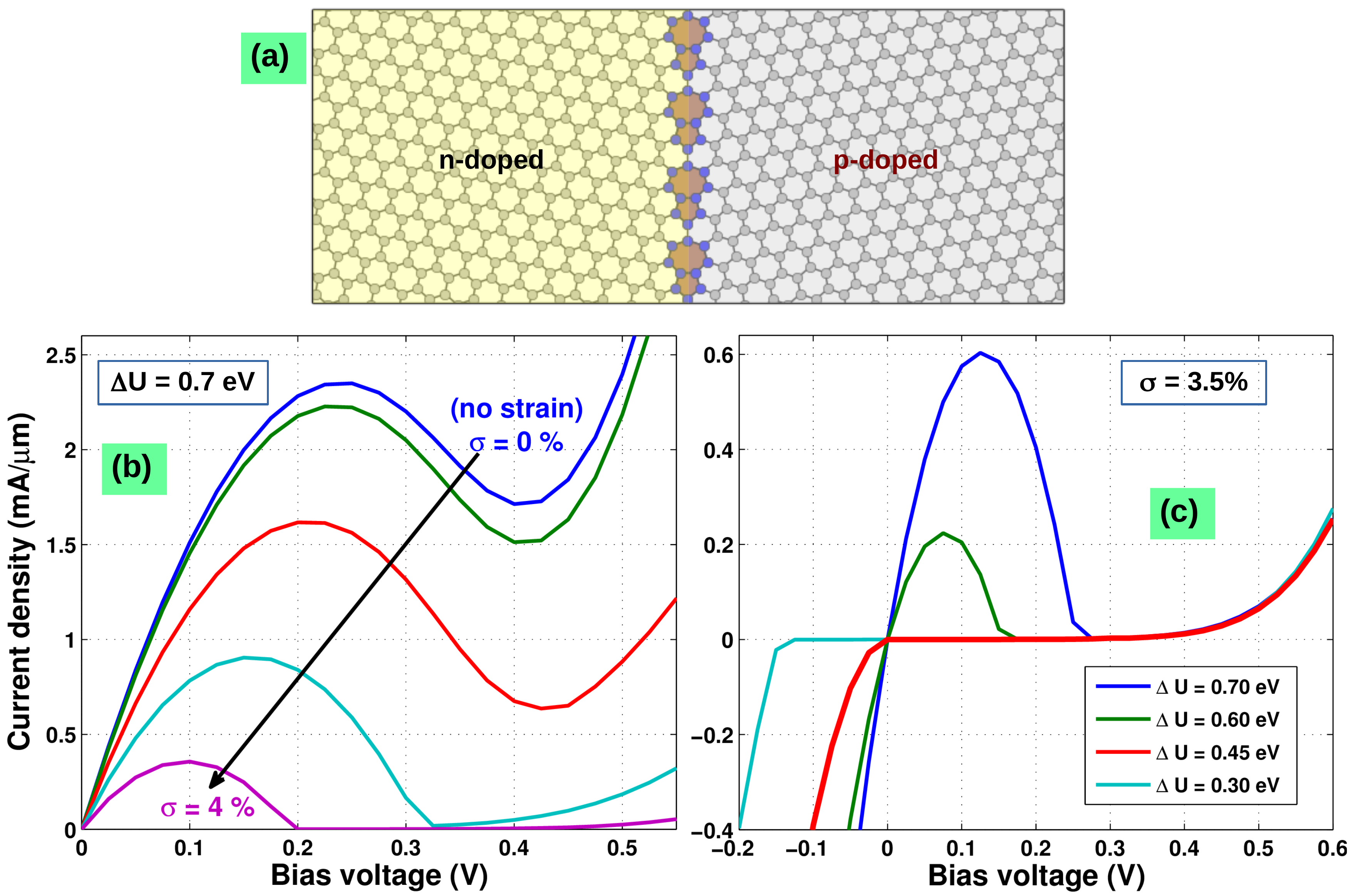}
	\caption{Schematic of the graphene p-n junction (a) and its corresponding I-V characteristics obtained for different applied strains (b) and different doping levels $\Delta U$ (c). The strain is applied in the direction $\theta = 45^\circ$ and all the simulations were performed at room temperature.}
	\label{fig_sim3}
\end{figure*}

Besides two applications mentioned above, this system can also be used in another electronic device, namely the p--n junction (or tunnel diode \cite{seab98}), where the interband tunneling between electron states of the n-side and hole states of the p-side is the main conduction mechanism. Our novel device, where one graphene domain is n--doped while the other one is p--doped, is schematized in Fig.11.a. The doping profile is characterized by the potential difference $\Delta U$ between two graphene domains that is additionally introduced in our tight-binding Hamiltonian shown above. Our idea consists in considering that in this device, while the strain can lead to the separation of Dirac cones of two graphene domains along the $k_y$-axis, the doping is used to generate their relative shift in energy. The combination of these two features can offer the possibility of modulating the energy gap of transmission function and hence generating a strong non-linear \textit{I--V} characteristics. In Figs.11.b-c, the \textit{I--V} characteristics obtained for different strains (b) and different doping levels (c) are displayed. For a fixed potential difference $\Delta U = 0.7$ eV (Fig.11.b), an \textit{I--V} characteristics with only a weak effect of negative differential conductance (NDC) is observed when no strain is applied. This is an expected result since the graphene system has a zero transport gap and the interband tunneling current can not be strongly switched off. However, when a strain is applied, though the peak current is reduced, the valley current can be very strongly suppressed, leading to a very strong NDC behavior with an extreme large peak-to-valley ratio. Indeed, it can reach values as large as $\sim$ 270 and 792 for strains of $3.5\%$ and $4\%$, respectively. Moreover, in Fig.11.c, the \textit{I--V} characteristics were calculated for different $\Delta U$ while the strain is fixed at 3.5$\%$. In a finite range of forward bias, the current is found to be strongly modulated by varying the doping level $\Delta U$. For an appropriate doping (i.e., $\Delta U \simeq 0.45$ eV here), the current can be totally suppressed in a finite range of forward bias. However, it still increases rapidly when applying and raising a reverse bias, implying a strong rectification effect in the device. 
\begin{figure*}[!t]
	\centering
	\includegraphics[width = 0.95\textwidth]{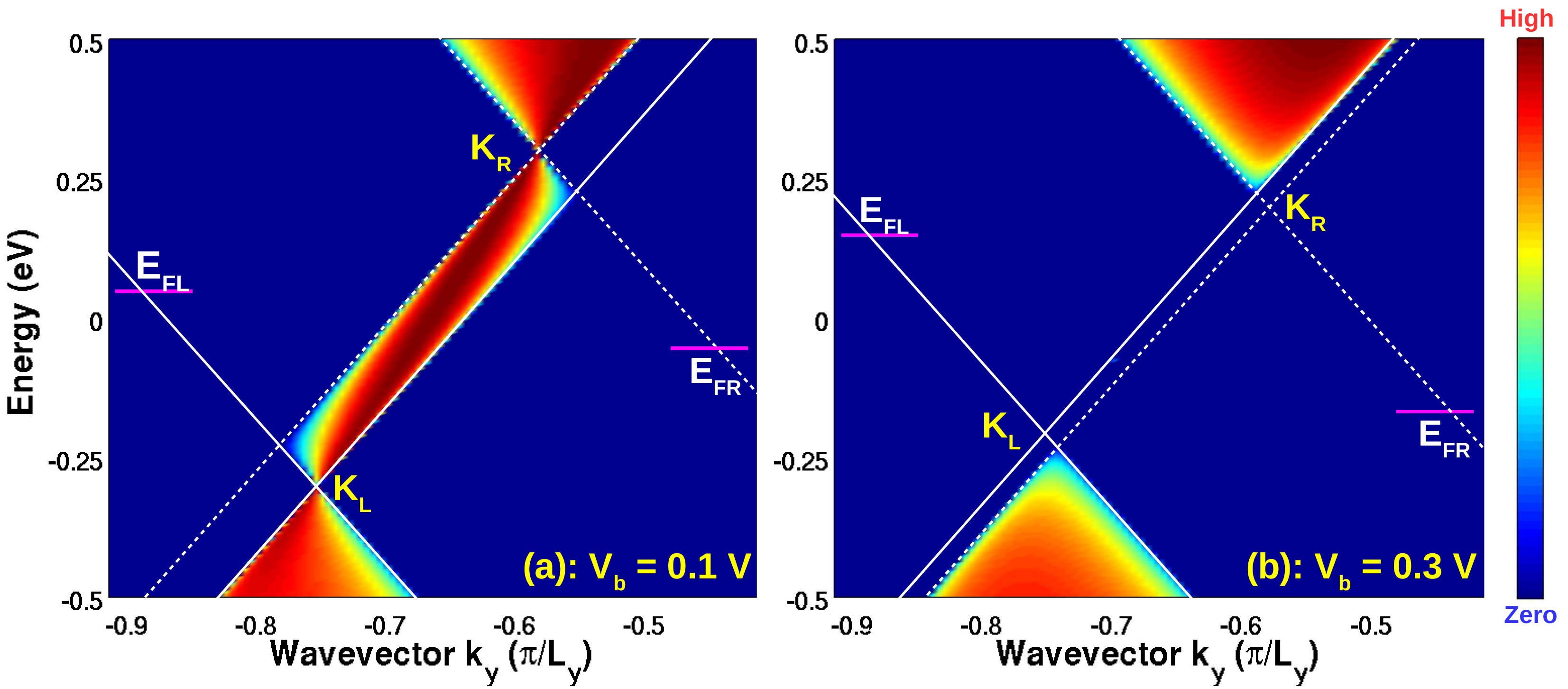}
	\caption{($E-k_y$)-maps of transmission probability in the graphene p--n junction in Fig.11 for different applied bias: $V_b$ = 0.1 $V$ (a) and 0.3 $V$ (b). The parameters $\Delta U = 0.7$ eV, $\sigma = 3.5\%$, and $\theta = 45^\circ$ are considered in the calculations.}
	\label{fig_sim3}
\end{figure*}

To explain the mechanisms related to these strong non-linear effects, ($E-k_y$)-maps of transmission probability obtained at different bias are plotted in Fig.12. The key ingredient consists in using the strain/doping-induced modulation of the band profile to engineer the energy gap and transmission through the graphene p--n junction. The transmission is large in the energy regions where the bands of the two graphene domains overlap, otherwise, an energy gap is achieved. At low bias and for high doping (i.e., $\Delta U > E_g$, where $E_g$ is the strain-induced gap obtained without doping), there are three overlapped regions (i.e., high transmission), namely, the high-energy (thermionic), middle (interband tunneling) and low-energy (thermionic) ones as shown in Fig.12.a. In the p--n junctions considered here, the transmission in the interband tunneling region contributes mostly to the current and a high peak-current is observed at low bias. When raising the bias, the potential profile is modulated and this interband tunneling region disappears at high bias (Fig.12.c). This leads to a large energy gap covering the energy window [$E_{FL},E_{FR}$] and hence the current is almost suppressed if a large strain is applied. Indeed, a very small valley current and a strong NDC behavior can be obtained as shown in Fig.11. Additionally, for a given strain (i.e., a given $E_g$) and a doping level with $\Delta U \simeq E_g$, the transport picture is as shown in Fig.12.b if a forward bias is applied and a current gap is hence obtained in a finite range of forward bias. If a reverse bias is applied, the transport picture is as shown in Fig.12.a and the current increases rapidly with the bias. This explains the strong rectification effect observed in Fig.11.b.

Thus, when considering the possibility of modulating the electronic properties of graphene using strain, the graphene GB systems appear to be very promising for some specific applications as in strain nanosensors, flexible transistors and p--n tunnel junctions with strong non-linear \textit{I--V} characteristics.

\section{Summary}

By means of atomistic simulations, we report on a theoretical study on the electronic transport in graphene systems containing a single grain boundary and provide a sytemmatical analysis on the effects of uniaxial strain and lattice symmetry on their electronic properties. First, it is shown that not only depending on the strain magnitude, the strain effects on the transport properties of the system are also strongly dependent on the applied direction and the lattice symmetry of the two graphene domains surrounding the boundary. On this basis, strain engineering is suggested to be used to open a finite transport gap in all systems where two graphene domains exhibit different orientations. By choosing appropriately the strain direction, a large transport gap (i.e., $\sim$ a few hundred meV) and/or a large modulation of the gap can be achieved with a strain as small as a few percents. The dependence of this transport gap on the lattice symmetry of graphene domains (i.e., their orientation, the shape and size of their supercell, their relative commensurability) is also carefully investigated and clarified. We additionally report on a peculiar phenomenon concerning  strain effects on the mechanisms of defect scattering in these graphene heterostructures. We show that in specific graphene grain boundaries where the energy bands have different parity symmetries (similarly to that in ZGNRs), strain effects can be used to reduce the GB defect scattering and significantly enhance the conductance. Finally, on the basis of a large strain-induced transport gap, we suggest that the graphene systems with a single grain boundary could be very promising for applications as highly sensitive strain sensors, flexible transistors and p--n junctions with a very strong non-linear \textit{I--V} characteristics.

Finally, note that an important factor may have an influence on our results. Since our simulations are performed on periodic grain boundary structures, the presence of disorders could have important impacts on their transport properties. In particular, the disorders break the periodicity of grain boundaries and thus can introduce some leakage current within the energy gap region. However, previous studies \cite{yazy10} have demonstrated that moderate disorders in grain boundaries cause only a low leakage current and hence maintain very low conductance within the transport gap. More interestingly, some efforts to control the periodicity of grain boundaries at the large scale have been experimentally realized \cite{chen14,yang14,luan15}. In particular, periodic grain boundaries as long as a few ten nanometers can be achieved after the thermal reconstruction of aperiodic ones \cite{yang14}. These samples of reconstructed polycrystalline graphene could be very useful to check the interesting findings predicted in this work for periodic GB graphene heterostructures.

\textbf{\textit{Acknowledgment.}} V.H.N. and J.-C.C. acknowledge financial support from the Fonds de la Recherche Scientifique de Belgique (F.R.S.-FNRS) through the research project (N$^\circ$ T.1077.15), from the Commmunaut\'{e} Wallonie-Bruxelles through the Action de Recherche Concert\'{e}e (ARC) on Graphene Nano-electromechanics (N$^\circ$ 11/16-037) and from the European ICT FET Flagship entitled "Graphene-Based Revolutions in ICT And Beyond" (N$^\circ$ 604391). This research in Hanoi is funded by Vietnam's National Foundation for Science and Technology Development (NAFOSTED) under grant N$^\circ$ 103.01-2014.24.

\end{document}